\documentclass[sigconf,screen,nonacm]{acmart}
\AtBeginDocument{%
  }

\setcopyright{rightsretained}
\copyrightyear{2025}
\acmYear{2018}
\acmDOI{XXXXXXX.XXXXXXX}
\acmConference[Conference acronym 'XX]{Make sure to enter the correct
  conference title from your rights confirmation email}{June 03--05,
  2018}{Woodstock, NY}
\acmISBN{978-1-4503-XXXX-X/2018/06}

\usepackage{xcolor}
\usepackage{xspace}
\usepackage{subcaption}

\usepackage{listings} %

\definecolor{keywordcolor}{rgb}{0,0,1}      %
\definecolor{stringcolor}{rgb}{0.58,0,0.82} %
\definecolor{commentcolor}{rgb}{0,0.5,0}    %
\definecolor{bgcolor}{rgb}{0.95,0.95,0.95}  %
\definecolor{numbercolor}{rgb}{0.8,0,0} %
\definecolor{keycolor}{rgb}{0,0,1} %

\lstdefinelanguage{json}{
    basicstyle=\ttfamily\footnotesize, %
    frame=single, %
    breaklines=true, %
    numbers=left, %
    numberstyle=\tiny\color{gray}, %
    showstringspaces=false,
    morestring=[b]", %
    morecomment=[l]{//}, %
    commentstyle=\color{gray}, %
    stringstyle=\color{stringcolor}, %
    moredelim=[l][\color{keycolor}]{\{} %
}

\lstdefinelanguage{swift}
{
  morekeywords={
    func,if,then,else,for,in,while,do,switch,case,default,where,break,continue,fallthrough,return,typealias,struct,class,enum,protocol,var,func,let,get,set,willSet,didSet,inout,init,deinit,extension,subscript,prefix,operator,infix,postfix,precedence,associativity,left,right,none,convenience,dynamic,final,lazy,mutating,nonmutating,optional,override,required,static,unowned,safe,weak,internal,
    private,public,is,as,self,unsafe,dynamicType,true,false,nil,Type,Protocol,
  },
  morecomment=[l]{//}, %
  morecomment=[s]{/*}{*/}, %
  morestring=[b]" %
}

\definecolor{keyword}{HTML}{BA2CA3}
\definecolor{string}{HTML}{D12F1B}
\definecolor{comment}{HTML}{008400}

\newcommand{\systemname}{A\textsc{thena}\xspace}

\begin{document}

\title{\systemname: Intermediate Representations for Iterative~Scaffolded~App~Generation~with~an~LLM}

\settopmatter{authorsperrow=4}

\author{Jazbo Beason}
\affiliation{%
  \institution{Apple}
  \city{Seattle}
  \state{Washington}
  \country{USA}
}
\email{jazbo@apple.com}

\author{Ruijia Cheng}
\affiliation{%
  \institution{Apple}
  \city{Seattle}
  \state{Washington}
  \country{USA}
}
\email{rcheng23@apple.com}

\author{Eldon Schoop}
\affiliation{%
  \institution{Apple}
  \city{Seattle}
  \state{Washington}
  \country{USA}
}
\email{eldon@apple.com}

\author{Jeffrey Nichols}
\affiliation{%
  \institution{Apple}
  \city{Seattle}
  \state{Washington}
  \country{USA}
}
\email{jwnichols@apple.com}

\renewcommand{\shortauthors}{Beason, Cheng, Schoop, and Nichols.}

\begin{abstract}
It is challenging to generate the code for a complete user interface using a Large Language Model (LLM). User interfaces are complex and their implementations often consist of multiple, inter-related files that together specify the contents of each screen, the navigation flows between the screens, and the data model used throughout the application. It is challenging to craft a single prompt for an LLM that contains enough detail to generate a complete user interface, and even then the result is frequently a single large and difficult to understand file that contains all of the generated screens. In this paper, we introduce \systemname, a prototype application generation environment that demonstrates how the use of shared intermediate representations, including an app storyboard, data model, and GUI skeletons, can help a developer work with an LLM in an iterative fashion to craft a complete user interface. These intermediate representations also scaffold the LLM's code generation process, producing organized and structured code in multiple files while limiting errors. We evaluated Athena with a user study that found 75\% of participants preferred our prototype over a typical chatbot-style baseline for prototyping apps.
\end{abstract}

\begin{teaserfigure}
  \centering
  \includegraphics[width=0.9\textwidth]{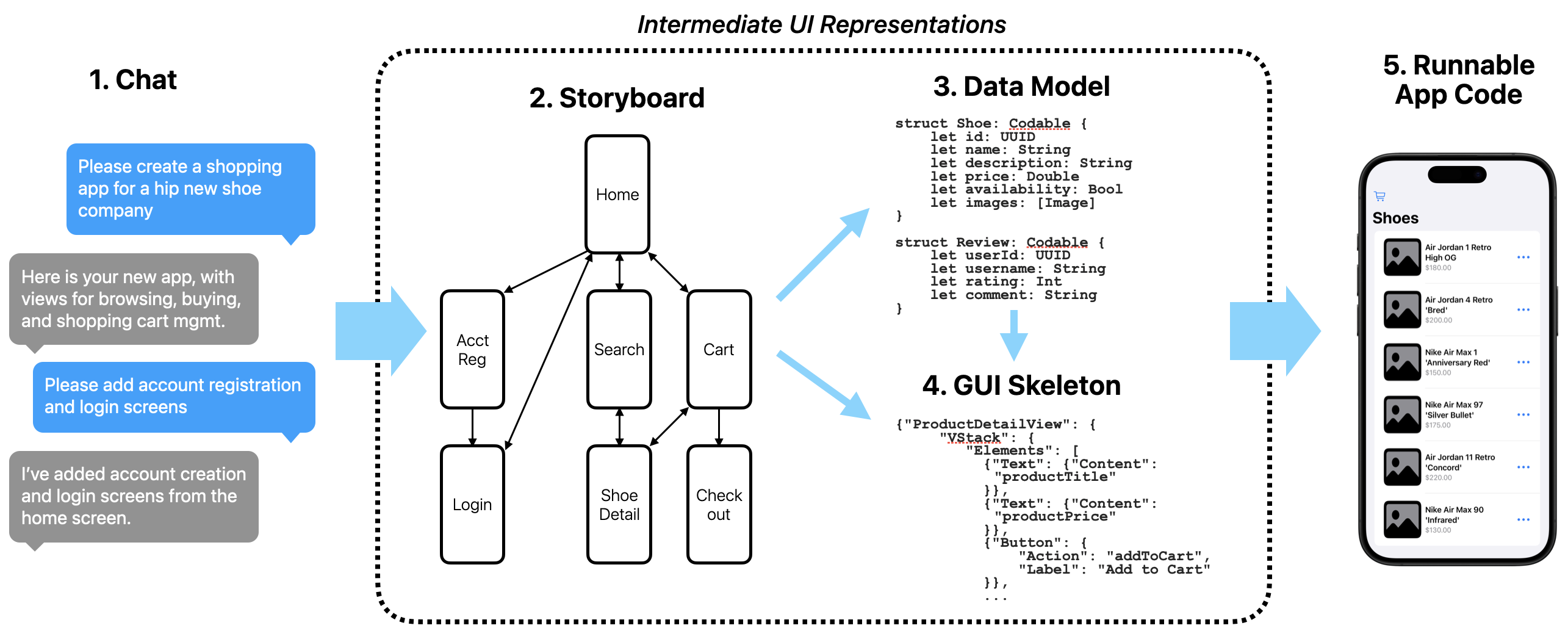}
  \caption{An overview of how our prototype system uses intermediate representations to scaffold the generation of an app with multiple inter-related screens. The primary interaction is through chat, in which the user and system can discuss the app, its structure, and UI elements. The chat is used to iteratively assemble three intermediate representations: a storyboard, a data model, and GUI skeletons for each screen. Ultimately, these intermediate representations are used to generate app code that can be run on a real device.}
  \Description{An overview of how our prototype system uses intermediate representations to scaffold the generation of an app with multiple inter-related screens. The primary interaction is through chat, in which the user and system can discuss the app, its structure, and UI elements. The chat is used to iteratively assemble three intermediate representations: a storyboard, a data model, and GUI skeletons for each screen. Ultimately, these intermediate representations are used to generate app code that can be run on a real device.}
  \label{fig:teaser}
\end{teaserfigure}

\maketitle

\section{Introduction}

Large Language Models (LLMs) have shown promise for accelerating the design stages of app development by helping retrieve~\cite{galleryDC, eldonTappability} and critique~\cite{peitongLLMHeuristics, peitongUICrit} UI designs.
Recent advancements have also improved LLMs' capabilities for generating code~\cite{humanEval}, conversational debugging~\cite{copilotConversation, chopraConversationalDebug}, and Q\&A to support code understanding~\cite{LLMUnderstandingCode}.
Modern LLMs nonetheless struggle to produce correct UI code due to its complex and specialized nature~\cite{wu-2024-uicoder}. Furthermore, LLM-based design tools generally only support designing for single views, and LLM-based code assistants target development scenarios that are unaware of the overall design intent of the app.

One challenge for LLM-driven UI code generation is that typical implementations produce code directly from user prompts, with only the conversation serving as context. All of the knowledge about the UI under development must be included in the code or the conversation, and all of the code is regenerated in response to each user prompt. Code may change radically at each iteration and important aspects of one iteration may be lost in the next. A potential solution is to create intermediate representations that describe the intended user interface at a higher level than code, use these as shared artifacts that can be referred to and modified in the conversation between the user and the system, and ultimately generate UI code based on these shared artifacts. In some ways, this approach is a modernization of previous work in model-based user interface development~\cite{puerta:mobi-d}, but in this case a conversation with an LLM drives the creation of the intermediate representations rather than tedious manual construction by designers and developers.
Our work also complements recent breakthroughs in software development agents which autonomously break down coding problems into solvable chunks, interact with developer tools, and iterate on their own code~\cite{dreamgarden, sweagent, copilotAgent}. We focus on using intermediate representations that reflect typical stages in app development to guide an LLM to generate multi-view apps.

In this paper, we explore three Intermediate Representations (IRs): a Storyboard that describes the structure of the screens in the app, GUI Skeletons for each screen, and a Data Model describing how the app stores and structures all of its data. These representations are all displayed in a unified editor, called \systemname, which also includes a chatbot-style conversational interface that allows the user to construct and iterate on these representations with the help of an LLM. Users can also edit the text of each intermediate representation. 

We chose these intermediate representations because we believe they reflect important aspects of the app development process, and can help users explore design possibilities and communicate their intent to the system. These representations are used to guide the generation of code that implements a finished app prototype in SwiftUI code~\cite{swiftUI}, which can be imported into the Xcode development environment for further editing. The generated code can be built in Xcode and the resulting binary will run natively on an iPhone.

We evaluate our approach's ability to assist developers with creating apps from scratch in a user study. In 25 minutes, all 12 participants were able to generate a complete, multi-screen app with navigation flows that could run on a real device. Compared to a baseline condition of a chat-based LLM interface, we found that developers using our prototype built functional apps with twice as many views and three times as much code in the same amount of time. Participants said that they would prefer to use Athena over the baseline if they were building a prototype from an initial app idea.

We also conducted a technical evaluation of our approach, in which we created ten representative apps and analyzed \systemname's output. Because of the larger number of views and greater amount of code, some bugs were present in \systemname's output. However, most apps contained only a few easily correctable bugs involving compilation or the navigation links between views.

It is important to note that we have only built a prototype to explore the use of intermediate representations in UI and app development, but is not a full app development environment on its own. For example, its unified editor does not support building or debugging code directly. There are other important aspects of app development that are not yet supported, such as connecting to external services over a network. Our purpose in this paper is to explore systems that can generate structured multi-screen app user interfaces through an iterative conversation between a user and an LLM. Other aspects of app development are left for future work.

Our paper makes the following contributions:

\begin{itemize}
    \item The design of 3 intermediate representations following common, critical steps in the UI development process: the Storyboard, a Data Model, and GUI Skeletons. These guide the code generation LLM while inviting comprehension and control by the human developer.
    \item Our prototype system with a chatbot-style UI, which integrates the above intermediate representations and generated code in a unified editor.
    \item A technical evaluation and user study demonstrating our approach's effectiveness in generating implementations of multi-screen app prototypes.
\end{itemize}

\section{Related Work}
\label{section:related-work}

Tools that assist with building applications have been a focus of HCI nearly since the beginning of the field, with work stretching back several decades. Early work focused on interface builders, which were direct manipulation tools that allowed a developer to assemble a user interface out of components while writing little or no code~\cite{myersRossonUIprogramming}. Other work introduced model-based UI development, which focused on generating applications from abstract descriptions of the intended functionality~\cite{szekelyModelBased}. Model-based UI development hoped to eliminate the need to write UI code, but unfortunately the models needed to define the UI were at least as tedious to create and maintain, and thus the work fell out of favor.

The storyboard concept, as used by \systemname, was explored in some of these early tools. For example, Denim is a sketch-driven web site builder that uses a storyboard as one of its fundamental abstractions~\cite{linDenim2000}. Highlight was mobile site re-authoring tool that also featured a storyboard as a fundamental aspect of its design~\cite{nicholsHighlight2008}. StoryDroid uses the storyboard concept to help with understanding existing apps through extracting activity transition graphs from Android apps~\cite{storydroid} for use as design inspiration.

\subsection{AI-Assisted UI Design}

More recently, but prior to the rise of LLMs, various AI techniques have been applied to assist in the design and implementation of user interfaces. Supple demonstrates how generating UI layouts can be formulated as an optimization problem~\cite{gajosSupple}. Other systems use content and style transfer~\cite{bricolage}, and editable layout constraints~\cite{amandaScout} to explore UI layout and design alternatives. All of this work focuses on single screens however, rather than multi-screen navigation, as we explore in \systemname.

Another set of work examines how to efficiently retrieve examples from large databases of user interfaces for design inspiration. Swire enables retrieval of UI designs from sketches~\cite{swire}. Gallery DC enables faceted search of UI components from design criteria~\cite{galleryDC}. Huang et al. applied deep learning methods to retrieve UI mockups from text or UI components~\cite{huangMockups2021}. These tools focus on single screens and do not generate app designs or code directly, but they might be useful for future integration into a tool like \systemname.

Large language models offer the ability to produce UI designs from natural language. Prior works use such models to generate low-fidelity~\cite{huangMockups2021} and medium-fidelity~\cite{feng2023designinglanguagewireframingui} mockups. These systems generate intermediate design artifacts for single screens that can be useful in early stage design work.
Diffusion models have also been used to turn rough wireframes into images resembling UI screens~\cite{weiUIDiffuser} and generate constrained UI layouts~\cite{forrestPlay}. These systems output images and vector drawings respectively, which can be useful as design motivation, but still require writing code to realize an implemented app. Other LLM-based systems also produce code, which will be discussed in the next section.

Finally, recent work has employed LLMs and multimodal models to evaluate user interfaces. Duan et al.~\cite{peitongLLMHeuristics} use LLMs to generate heuristic evaluations of UI mockups.
Single screen evaluation was also explored in UIClip~\cite{wuUIClip}, which is a CLIP-based model fine-tuned for quality and relevance assessment given pairs of screenshots and descriptions.

\subsection{Generating UI Code}

Large language models and other ML techniques have been used to generate code, either from natural language or more structured information. One example is Relay, a Figma plugin that exports UI model code from structured mockups, which helps smooth the handoff from designers to developers~\cite{relay}. Other systems, such as ReDraw, use ML techniques to retrieve and assemble UI code~\cite{moranUIPrototyping}.

Other systems start from image inputs, such as screenshots, and generate UI code directly. pix2code~\cite{pix2code} was one of the first systems to do so. Chen et al.~\cite{chunyangUISkeleton} developed a system that infers a \textit{GUI skeleton}, similar to that in \systemname, from an image of a Android UI. This skeleton can then be used by a developer to implement the UI in code. Icon2Code operates at a finer level of granularity, where it suggests callback code for a given icon on a particular screen~\cite{icon2code}. DCGen refines the process, by dividing screenshots of webpages into segments, generating code for each segment, and then assembling that code together~\cite{dcgen2024}. These systems are limited to generating code for individual screens, without the shared context or artifacts used to style multi-view apps.

Other approaches have focused on fine-tuning open source LLMs for generating UI code specifically. UICoder is one example, which introduces a self-supervised learning technique that enables small, open-source LLMs to generate SwiftUI code for individual screens~\cite{wu-2024-uicoder}.

\section{Building Apps with \systemname}

To illustrate how \systemname can be used to create SwiftUI apps through progressive revision of intermediate steps, we present the following scenario.

Missy is an independent iOS developer based in Jamaica, where traditional banking often falls short of meeting local community needs. Like many in her community, Missy’s mom, a small business owner, depends on informal communal finance systems known as partnerships to manage and move money effectively. A partnership works by pooling fixed weekly contributions from each member, with a rotating payout awarded weekly to a different person until everyone has received their payout. Growing up, Missy often saw how informal record-keeping and casual agreements could lead to confusion, misunderstandings, or disputes over who paid what, and when each participant would receive their payout. Determined to help her mom and neighbors, Missy envisions a straightforward yet powerful app to address these challenges.

Eager to rapidly prototype her idea, Missy launches \systemname. She types a description capturing the essential features and navigation flow of her app (\autoref{fig:teaser}):

\small
\begin{quote}
    ``Please create a finance management app tailored specifically for informal community savings groups (partnerships) in Jamaica. It should allow users to easily track weekly payments, clearly highlight outstanding contributions, and visually display the rotation order for payouts. Let’s start with login/sign up flows, payout schedules, participant management, and payment tracking.''
\end{quote}
\normalsize

\systemname generates an Storyboard with a reasonable set of screens, but lacking a sign up screen, so Missy requests, \textit{``Please add sign up to the flow.''}

\systemname adds the sign up screen to the storyboard while keeping the other screens in place. Satisfied with these adjustments, she generates the secondary intermediate representations: GUI Skeletons and Data Model. 

While inspecting how payments are represented in Data Model, she recognizes that billing addresses are missing. She asks \systemname to add the field, which leads to a change in the screen's GUI Skeleton.

Now that the intermediate representations meet her expectations, she generates the SwiftUI views and exports them to an Xcode project for closer inspection. In Xcode previews, Missy identifies a styling improvement for the DrawScheduleView. Back in \systemname, she provides detailed guidance back to \systemname:

\small
\begin{quote}
    ``Please style the payout schedule view nicely. Add a card for each row with an image for the person's profile pic. Stub it out for now with sf symbol. Also, I wanna be able to see the person's name, the date of the payout, their profile pic, and the amount of the payout in each row in this view.''
\end{quote}
\normalsize

\systemname generates a modification plan for all the changes in her request, modifies the intermediate representations, and updates the SwiftUI code. Missy exports again, for another look at her progress.

In less than an hour, her prototype is complete far faster than she imagined possible. After a brief period of further refinement, adding a backend and some custom styling, she confidently submits the final product to publish her app.

\section{System Implementation}

Our prototype of \systemname is a macOS desktop app written in SwiftUI. The application incorporates a chat interface, lightweight text editors to modify text-based intermediate representations (IRs), and a Storyboard IR renderer that draws apps' navigation graphs using Graphviz. The screenshots of the \systemname user interface can be found in the Appendix. 

\systemname operates in three phases: 

\begin{enumerate}
    \item an initial generation phase based on the first message,
    \item the main phase in which the user interacts with the system and IRs are created and updated at each step, and
    \item a code generation phase where code is generated based on the content of the IRs.
\end{enumerate}

The current implementation of \systemname separates the code generation phase from the main phase due to the latency required to generate code.
After generating code, the user may return to the main phase and continue to modify the IRs. However, any changes made to the generated code in Xcode cannot currently be reflected back into IRs.

\subsection{Intermediate Representations}
\label{sec:system-ir}

\systemname maintains three types of IRs, which are used as building blocks to scaffold code generation: a \emph{Storyboard, Data Model,} and multiple \emph{GUI Skeletons}. We chose these IRs since they represent common incremental design milestones developers encounter when implementing apps: identifying the different views an app should include, the data manipulated between the views, and how the views are rendered.

\subsubsection{Storyboard}
The Storyboard defines the structure of the app, represented as a directed graph with each screen as a node and the navigation paths between screens as edges. The graph is represented in JSON with each node having an ID, name, description, and a list of outgoing edges with IDs of other nodes in the storyboard. Each node also has a SwiftUI view name, which is used during code generation to ensure consistent naming across multiple source files.

\begin{figure}
    \centering
    \addtocounter{figure}{1}
    \includegraphics[width=0.6\linewidth]{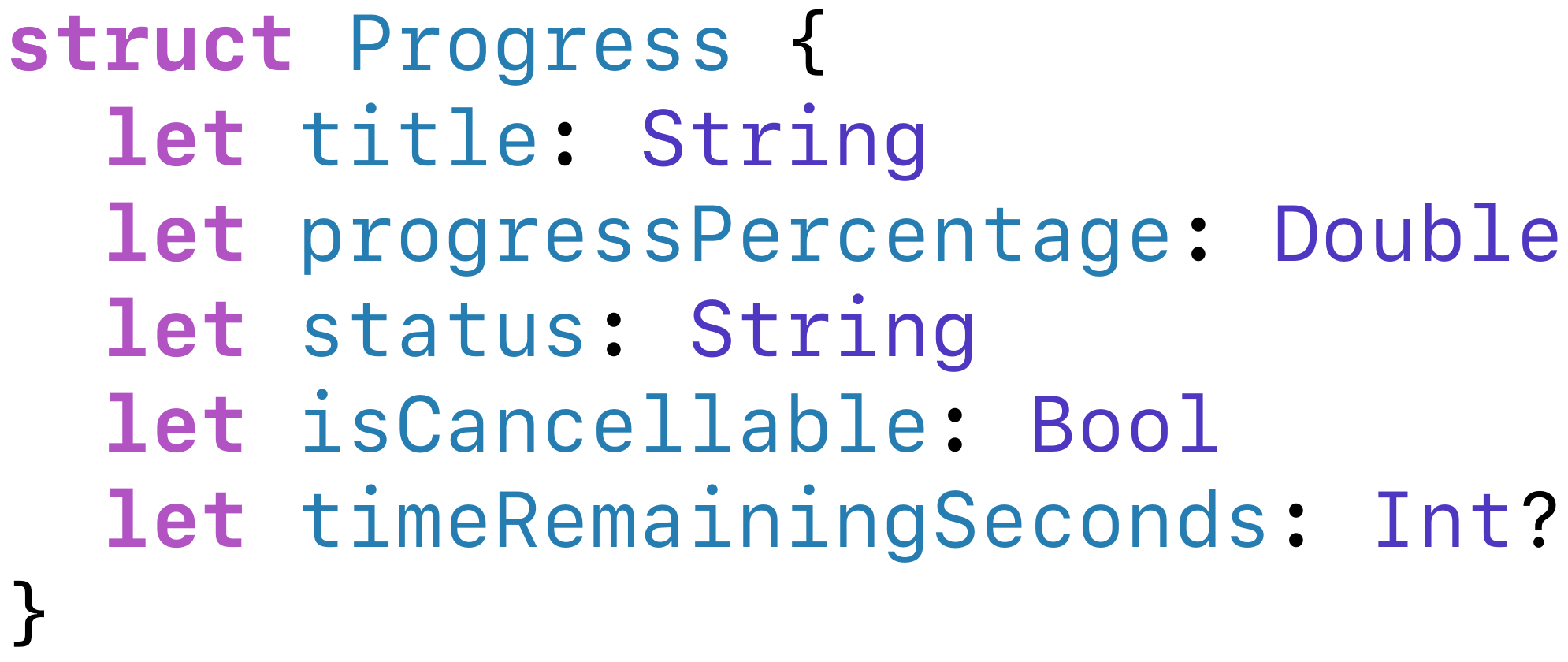}
    \caption{A data entity for a progress bar view with a title, percentage completion, and other data fields.}
    \label{fig:system.progressentity}
\end{figure}

\subsubsection{Data Model}
The Data Model contains the types of data to be displayed or manipulated by the app (e.g., a book or a restaurant reservation), and is made up of a set of data entities. Each entity is represented as a struct in the Swift programming language and may be shared across multiple Storyboard screens. In practice, we found most entity fields generated by \systemname use Swift primitive types, but nested entities and more complex types are not uncommon.
In our current design, some entities reflect data that might be stored in a database or be populated from a web-based data source, while other entities represent ephemeral data needed to operate a view, such as the \textit{Progress} entity shown in \autoref{fig:system.progressentity}.

\begin{figure}
    \centering
    \includegraphics[width=0.8\linewidth]{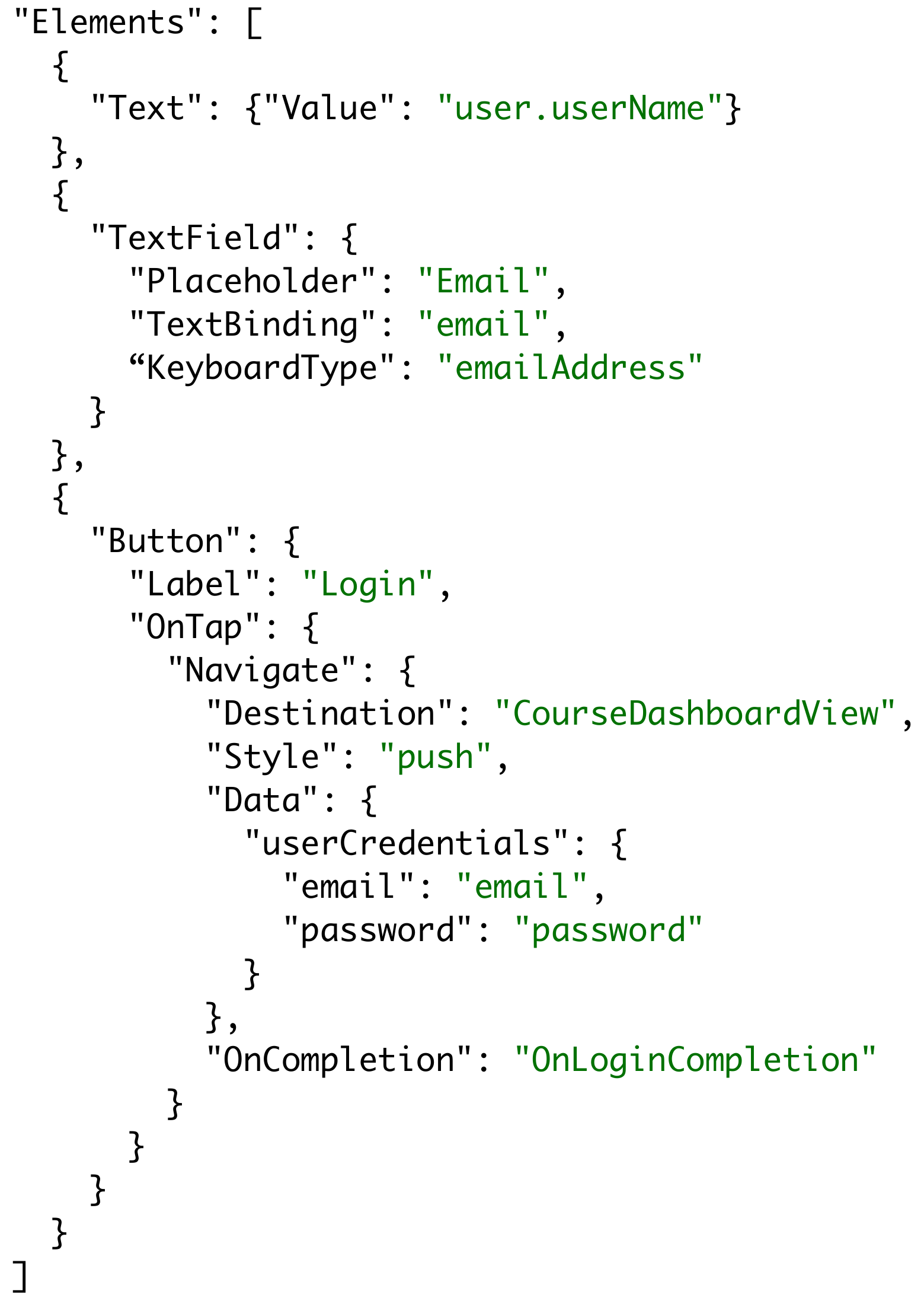}

    \caption{An abbreviated GUI skeleton of a password entry screen
    where the user name was entered on a previous screen but is
    displayed on the current screen.}
    \label{fig:system.loginskeleton}
\end{figure}

\subsubsection{GUI Skeleton}
A GUI Skeleton specifies the content and interactivity of a screen. Each screen present in the storyboard has an associated GUI skeleton, and is represented as SwiftUI pseudocode that specifies navigational and structural attributes (e.g., NavigationView and VStack) along with basic UI elements such as Button and Text. Action handlers, for example on buttons, are represented with strings describing the action to be taken, but not code needed to actually execute the behavior. Skeletons can also refer to data entities and their fields to populate UI elements. An example of a simple login screen can be seen in \autoref{fig:system.loginskeleton}.

The SwiftUI view name in the Storyboard, the use of Swift structs for the Data Model and the use of SwiftUI pseudocode for the GUI Skeleton were all chosen through empirical testing to ensure higher quality code generation. In particular, the use of Swift code and pseudocode representations ensures that less translation is required from the IRs to the final code, and guarantees that modifications to the IRs will be representable in Swift and SwiftUI later in the process.

Note also that these representations have dependencies on each other. For example, the Storyboard defines each of the screens that are present in the app and a GUI Skeleton is generated for each screen. As the Storyboard is updated, the GUI Skeletons must be updated as well. The GUI Skeletons are also dependent on the Data Model and must only name entities and fields that exist. These dependencies affect the order in which the IRs are modified, as we will see in the next section.

\subsection{LLM Modification of IRs}
\label{sec:llm-modifications}

When a user provides a chat message or directly edits the text content of IRs in \systemname, the changes are sent to a planning prompt which determines what changes need to be made to all IRs. This approach is related to common decomposed prompting techniques~\cite{decomposedPrompting, planAndSolvePrompting} used by other agentic systems~\cite{dreamgarden}.
Specifically, the planner is prompted to decompose the user's request (or provided changes) into atomic operations to create, update, or delete Storyboard views, connections between Storyboard views, Data Model entities, and GUI Skeleton files.
Small requests, such as adding or removing a button from a view, may only require modification of that view's GUI Skeleton, whereas larger requests may modify multiple intermediate representations.

\begin{figure}
    \centering
    \includegraphics[width=\linewidth]{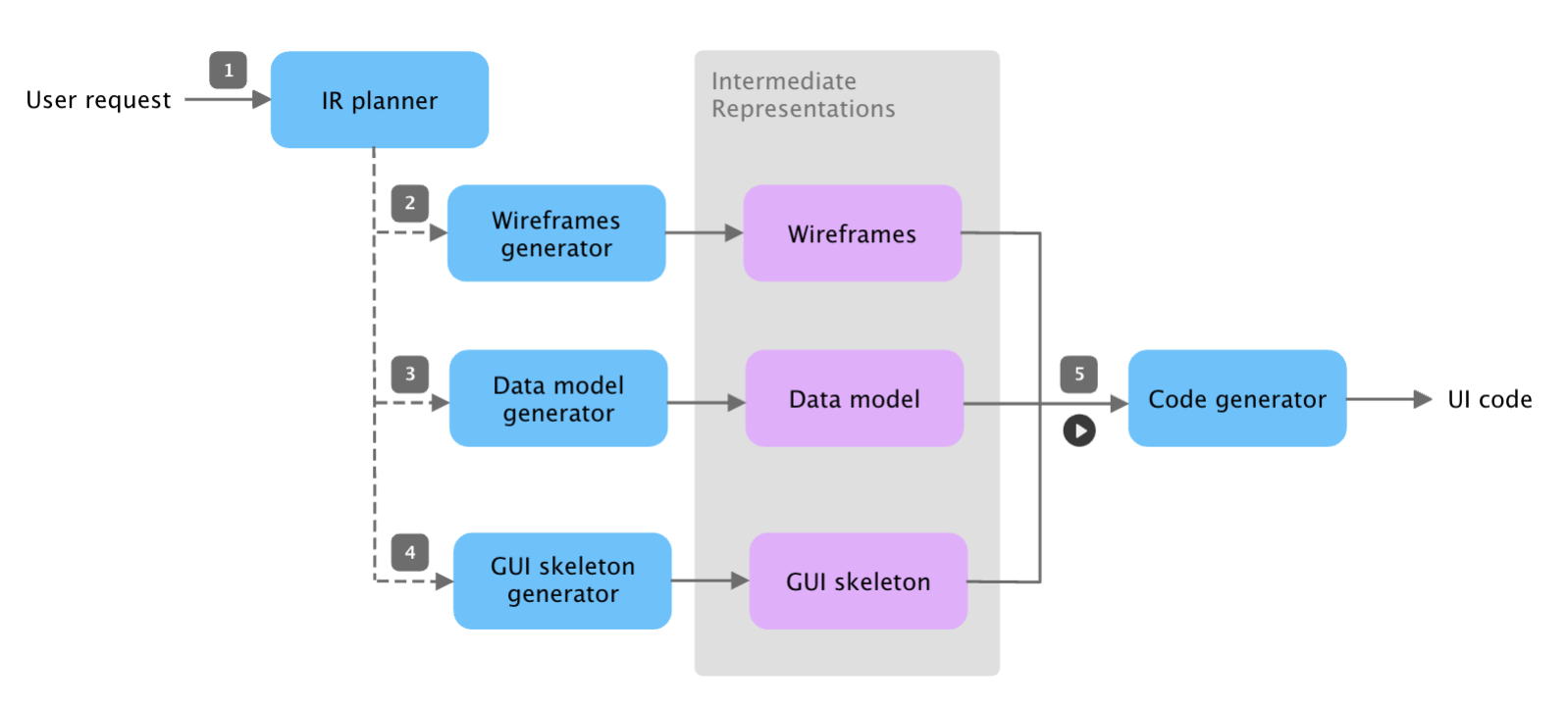}
    \caption{The prompt flow of \systemname system for modifying the IRs. A user message first triggers a plan prompt, which decides which IRs must be updated. Changes are then made as needed, starting with the Storyboard, then Data Model, and finally the GUI Skeletons. Separately, code is generated from the same IRs.}
    \label{fig:system.promptflow}
\end{figure}

After the plan is produced, the IRs are each modified by an LLM in a cascading order that takes into account the dependencies between the representations. The Storyboard is modified first, followed by the Data Model, and then any affected GUI Skeletons (\autoref{fig:system.promptflow}).

Storyboard modification is done by prompting an LLM with the current Storyboard and the changes requested by the plan. Additions, removals, and changes in edges are all processed in a single LLM call, but in the future removals and edge changes might be implemented by tools that change the internal data structure directly without requiring an LLM. Additions would likely always require an LLM to fill in a name and description, and ensure edges across the Storyboard are appropriately modified to include the new addition.

Data Model modification follows Storyboard modification and is implemented in a similar way. For each change specified by the plan, an LLM is prompted with the current Storyboard, current Data Model, and the requested change. The LLM returns an updated Data Model. All changes are again processed in one LLM call, but in the future, some changes could likely be processed by a tool without requiring an LLM.

GUI Skeleton modification follows Storyboard and Data Model changes, as the skeletons are dependent on both. For each change specified by the plan, an LLM is prompted with the current Storyboard, Data Model, GUI Skeleton and the requested change, and returns a modified skeleton. As each GUI Skeleton is independent of the others, these changes are processed with parallel LLM calls to speed up processing.

Once the IRs are up-to-date, it is possible to generate complete SwiftUI code for the app.

\subsection{Code Generation}
\label{sec:system-codegen}

A single LLM call is used to generate all of the necessary code, which takes all of the IRs as input, and produces all of the output files encoded in a JSON format that can be parsed and placed into a typical SwiftUI file structure. Before settling on the single call approach, we tried several other approaches, including generating all views simultaneously from separate LLM calls, and ordering the view generation using a post-order traversal of the views so that child views were generated first and parent views were generated with knowledge of their children’s code. Both resulted in more errors than the single call approach.

Through experimentation, we found that the quality of UI code generation could be notably improved by also providing a ``design scaffold'' to the code generation call which describes the colors, fonts, navigation and other aesthetic and structural attributes to use while producing the code. The scaffold is not currently available to or modifiable by the user, though in the future we intend make it possible for the user to directly edit the scaffold and also modify it through the messaging system. To increase design variation in our current system, the design scaffold is generated by an LLM using a prompt from the initial query made by the user.

After the code is generated, it can be imported into an Xcode project, built, and run on a local or simulated iPhone device.

\subsection{Initial Generation}
\label{sec:system-initial-gen}

When the user first interacts with \systemname, there are not yet any existing IRs, so it is necessary to create a first version of each from scratch based on the user’s initial message. In general, we use the same prompts as described above, but we found that when creating a large number of GUI Skeletons from scratch, navigation capabilities would be left out or not fully formed. For this reason, during the initial generation step, we ask an LLM to create a ``navigation plan'' that describes the overall navigation design for the app based on the initial generated Storyboard. This plan is provided as an input to the creation of each initial GUI Skeleton. After the initial GUI Skeletons are constructed, we discard the plan and it is not used subsequently, even when new GUI Skeletons are created later. It is future work to understand whether there is value in retaining the plan or making it user inspectable or modifiable.

\section{User Study}

To investigate how the intermediate representations used by \systemname facilitate application prototyping, we conducted a user study with 12 iOS developers.

\subsection{Methods}
\subsubsection{Study design and procedures}
Our user study follows a within subject design which includes two conditions: the \emph{\systemname} condition, where participants prototype an iOS app using \systemname, and the \emph{baseline} condition, where participants prototype using ChatGPT's GPT-4o model, the same underlying model used by \systemname. The baseline provides similar chat and code generation capabilities to \systemname, but lacks the intermediate representations.
The study includes two open-ended app development tasks: creating an app for language learning through songs (\emph{Music Task}), and an app for renting parking spaces (\emph{Parking Task}). Participants conduct one task in the baseline condition and the other task in the \systemname condition, counterbalancing conditions and tasks to mitigate ordering effects.

Participants started the study with sharing previous experience and challenges with designing and developing iOS apps. They were then asked to complete the two app development tasks, each within 25 minutes. A pre-recorded demo video was shown before the \systemname condition for onboarding. For each task, participants were instructed to imagine that they were an independent developer and that the goal was to implement a low-fidelity prototype of a multi-screen app to demonstrate basic ideas for each view.
Once the participant was satisfied with the generated code, they were instructed to export the code into Xcode, fix any bugs, and complete the application UI.

During the tasks, participants were encouraged to think-aloud and we occasionally probed for their experience with the IRs. After each task, participants were asked to reflect and answer a series of Likert scale and comparison questions.
After both tasks, they were asked to indicate their preferences between \systemname and the baseline for designing navigation flow, developing prototypes, and generating code for real-life app development scenarios. Detailed questions are shared in the Appendix. Finally, participants were prompted to give any feedback on improvement or propose any new features to \systemname. 

\subsubsection{Participants}
We recruited 12 participants (3 females, 9 males) from a large technology company in the United States via internal messaging channels that cover a broad set of teams, roles, and experience levels across the organization. All participants had experience with iOS app development using the Swift programming language (minimum experience: 8 months, maximum experience: 18 years), with various job titles including software engineer, research engineer, and research scientist, from various departments in the company. All participants had experience using generative AI in programming tasks. Participants' profiles are detailed in Appendix. All participants received a \$15 gift card as compensation.

\subsubsection{Data collection and analysis}

We collected audio and video recordings of the study sessions with the consent of the participants and transcribed them into text. 
We followed a thematic analysis approach \cite{guest2012applied} to analyze the qualitative data. 
For rating and comparison questions, we report the descriptive statistics.

\subsection{User Study Results}

\begin{figure}
    \centering
    \includegraphics[width=\linewidth]{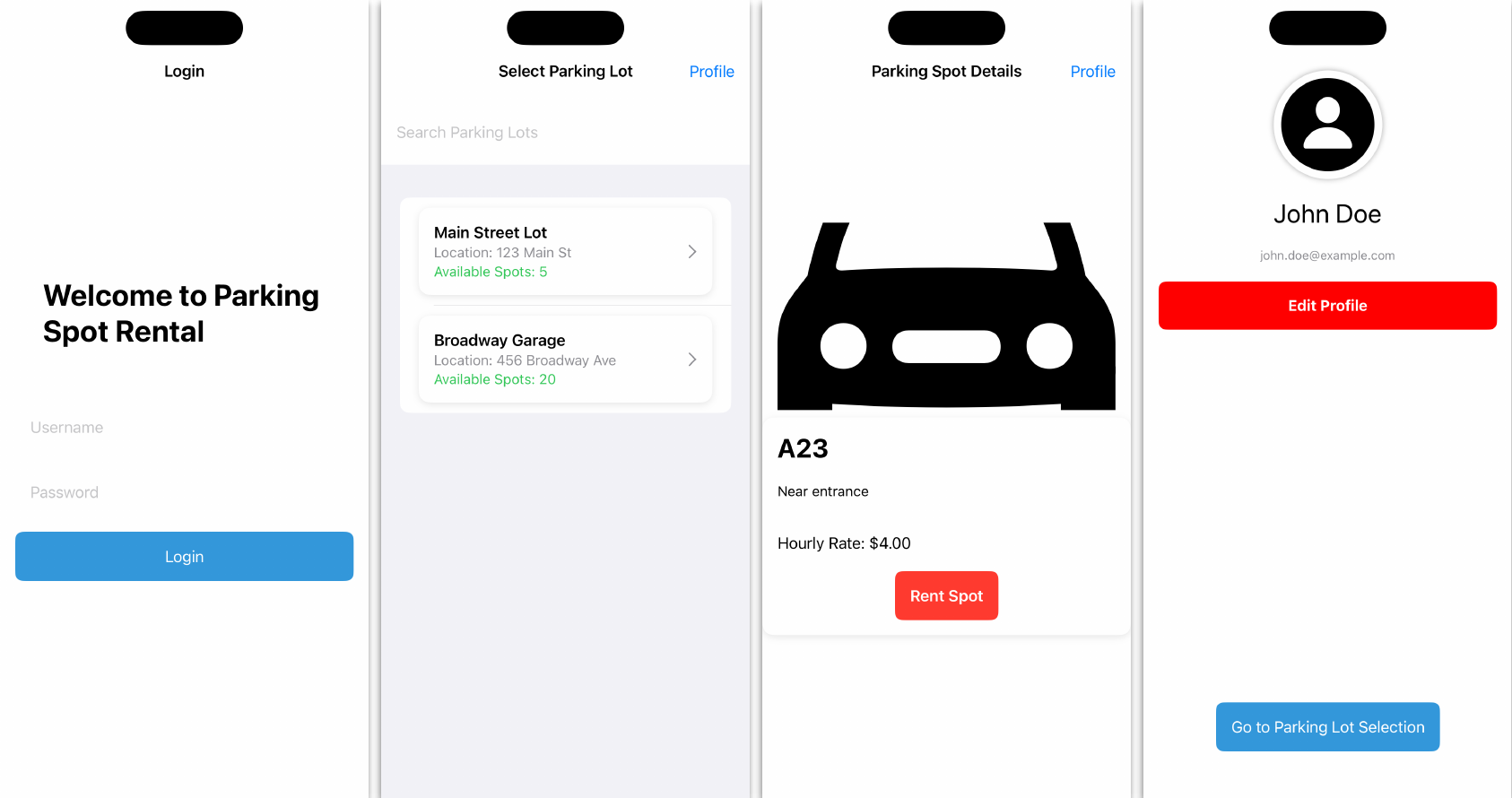}
    \caption{Screens from the app that P11 created during the user study with \systemname.}
    \label{fig:studyScreenshots}
\end{figure}

Overall, participants appreciated \systemname's ability to support developers in building iOS application prototypes with multiple screens. For tasks using \systemname, all participants were able to generate code for their prototypes in the 25-minute task time. 
\autoref{fig:studyScreenshots} shows a prototype that P11 made during the Parking Task using \systemname.
Participants shared their excitement for \systemname, noting that \systemname generated prototypes that matched their expectations. 

\begin{figure}[h]
    \centering
    \begin{subfigure}{0.45\textwidth}
        \centering
        \includegraphics[width=\linewidth]{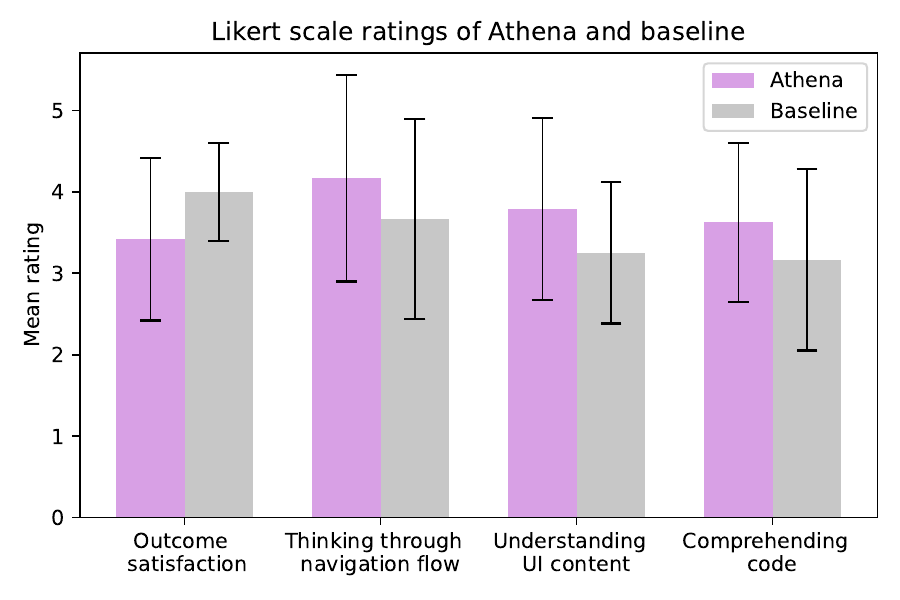}
        \caption{Likert scale ratings}
        \label{fig:likertResults}
    \end{subfigure}
    \hfill
    \begin{subfigure}{0.5\textwidth}
        \centering
        \includegraphics[width=\linewidth]{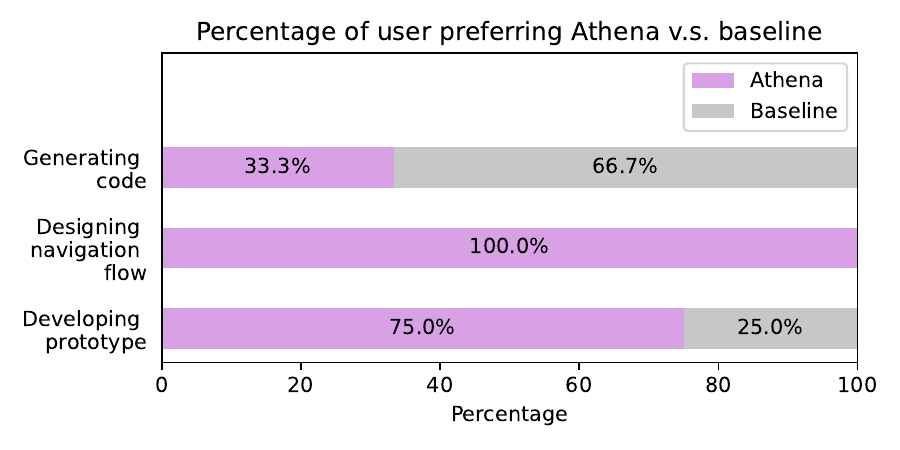}
        \caption{Percentage of user preferences}
        \label{fig:comparisonResults}
    \end{subfigure}
    \caption{User evaluation results}
\end{figure}

\subsubsection{\systemname is preferred in early stages of prototyping}
Participants expressed a desire to use \systemname early in the prototyping process. All 12 participants unilaterally preferred \systemname over the baseline for prototyping navigation flows. 9 of 12 (75\% of) participants said they would prefer to use \systemname over the baseline if they were building a prototype from an initial app idea (\autoref{fig:comparisonResults}).

\subsubsection{IRs matched participants' mental models and improved understanding of generated code}
Participants shared in qualitative feedback that \systemname's IRs matched the way they approached app prototyping, where they would first sketch out the navigation flow, then move on to the specific parts of the app:
``[\textsc{Athena}] really works like how I would build an app. And it's nice, like the idea that you can iterate on just the navigation flow before moving on to any other step'' (P2); 
``I really liked the idea of being able to iterate through the concept almost like formulating the idea before putting it in Xcode,'' (P4).
P5 shared that if they were to start something new, they would ``try to work as much as I can in \systemname to before I hit the build.''

In addition, some found the IRs helped break down code generation into steps, which allowed them to understand, iterate, and verify the generated code: ``if the interaction was that I gave it prompt and then I get an Xcode project out of it, then I would need to understand like all of it at once. Whereas by breaking it down into just the data model and then just the GUI skeleton, which at least gives me a sense for how a screen is going to be laid out, now I can be more confident when I see the Xcode project, what actually happened...it's sort of like showing it to me in steps.'' (P11)

\subsubsection{\systemname supports generation of complex applications}
The prototypes that participants created using \systemname were more complex than that of the baseline. On average, the prototypes created with \systemname contained 6.0 (SD = 2.2) views and 353.9 (SD = 92.8) lines of code, compared to 3.1 (SD = 1.1) and 117.8 (SD = 36.8) of the baseline. Details of apps that each participant generated using \systemname and the baseline can be found in the Appendix.
Participants also mentioned they would prefer to use \systemname over the baseline if they were working on a complex application involving multiple views: ``I think Athena's strength is it gives you this multiple navigational views... If you want to design a more complete app with like a login screen, like a welcome screen or like maybe just something to select which city you are in, [the baseline] is kind of blocked on what to do.'' (P1)

\subsubsection{Complexity of generated applications led to more LLM coding errors and latency, impacting usability}
As a result of \systemname allowing for more complex app structures and generating multiple code files, the prototypes created with \systemname contained more bugs (mean = 10.4, SD = 13.4) compared to that of the baseline (mean =  1.25, SD = 3.1). However, most bugs produced by \systemname in the user study were easy to fix for both novice and experienced iOS developers. All but one participant (P2) were able to fix all bugs in the generated code and successfully run a functional prototype by the end of the task. We analyze the bugs that \systemname may generate in more detail in \autoref{sec:technical_eval}.

The overall participant satisfaction with \systemname is $3.42$ (SD = $1.00$), which is between ``neutral'' and ``satisfied,'' compared to $4.00$ (SD = $0.60$) with baseline (\autoref{fig:likertResults}). From the qualitative feedback, we learned that the main reason the satisfaction score of \systemname is lower than the baseline is the latency in \systemname's generation (sometimes on the order of 1 minute to generate IRs) and the fact that the generated code contained bugs that required resolution.

In the following sections, we unpack how the IRs supported key aspects in prototyping the UI of an application.

\subsubsection{Storyboard}

The Storyboard generated by \systemname offered a useful scaffold for participants to think through the screens of the application at an early stage, and promoted ideation for navigation flows.
Participants shared that \systemname helped them think about navigation flows in ways they had not thought about before (rating their agreement as $4.17$ (SD = $1.27$), between ``agree'' and ``strongly agree''), compared to $3.67$ (SD = $1.23$) when rating the same statement about the baseline. 

\paragraph{Mapping out the full picture at the beginning}
Participants shared that the Storyboard generated by \systemname provided them with an effective visual representation of the navigation flow in their prototypes.
``It's like a nice visual summary of all the views, which is something that I think would be missing from [the] interface of just code... It's forcing me to do a more top down approach to design and development, that's not necessarily really a bad thing.'' (P8)
The Storyboard also helped show the scope of the app in the beginning of the development workflow, which could enable developers to arrive at a more complete design ahead of implementation. P2 shared their experience developing an app outside the study that they initially thought would be simple, but later realized that a more complex user flow was needed and had to modify the codebase significantly. P2 reflected that if they could have used \systemname, the Storyboard could help them ``understand the sheer scale of the app that [they] were undertaking.''
In contrast, with the baseline, participants saw code immediately, which could lead to premature focus on implementation details. 
P5 shared: ``with ChatGPT it could be not like what you think it is or how you interpreted what it spat out. [\textsc{Athena}] makes me kind of be on the same page on the storyboard.''

\paragraph{Encouraging early-stage exploration and iteration}
The visual representation of the screen flow also spurred iteration and ideation at the beginning of the prototyping process.
Seeing the Storyboard at such an early stage also encouraged debugging and iteration: ``I was able to spot without having to read through a bunch of code...ok, this is not what I was thinking in my head, so that definitely led to a much faster iteration approach as opposed to generate and read the code.'' (P8)
In another example, P5 found the entry point in the initially generated storyboard different from what they had imagined, so they prompted \systemname to make the home view as the entry point, and were able to quickly verify that the change was applied correctly.

The Storyboard also gave users novel design ideas.
When completing the Music Task, P2 got two screens from the Storyboard that \systemname generated that were not part of their initial imagination: ``the idea of grammar versus vocabulary is not something that occurred to me at all, but actually I think is a really good idea of kind of seeing it on paper.''

\paragraph{Suggestions for Storyboard}
Participants brought up two major suggestions for the future development of Storyboard. First, the Storyboard could include more details on the UI design of each screen, such as low fidelity wireframes of UI components. In P3's words, such details could ``entirely captur[e] the prototyping experience I want.'' Second,  participants hoped to directly manipulate the screens and navigation arrows in the storyboard, as opposed to describing the changes that they wanted in text-based chat with the LLM.
P4 found it would be more straightforward for them to directly edit the Storyboard for fine-grained iterations on navigation between views: 
``the arrows are probably something that I would want to draw myself, or at least rearrange myself, and, given a particular GUI, I think being able to create a screen and then connect it to the flow would be useful.'' (P4)

\subsubsection{Data Model}
\paragraph{Scaffolding understanding of UI content on each view}
On average, participants rated $3.79$ (SD = $1.12$) when asked about their agreement with the statement that they think \systemname helped them understand the UI content of the app to be generated (between ``neutral'' and ``agree''), compared to $3.25$ (SD = $0.87$) when asked about the baseline.
Participants highlighted that the Data Model helped them discover possible UI elements to include that they might have overlooked in their design (P12).

\paragraph{Providing a starting point for the implementation}
Participants shared that creating data in the app is typically a tedious process. The generated Data Model helps reduce this effort by offering a boilerplate of data that will be presented in the app:
``All that setup of just getting some views going, just [to] get some Data Model going, that takes like three hours of my time for an app that's complicated... But I need to start with something. So, yeah, the model's kind of doing my job for me, getting rid of like all the boilerplate set up, that's super handy.'' (P7)
Having the Data Model as an independent IR also helped participants focus on the UI content: 
``I really like the view where I'm able to see just the Data Model, nothing else, so I could really focus on what the Swift interpretation of what I'm trying to create. I think that's very important and I think having a lot of other things in the way could be very distracting.'' (P5)

\paragraph{Suggestions for Data Model}
Although participants acknowledged the functionality of the Data Model, some preferred to generate code directly after getting a satisfying Storyboard. Some participants had a hard time imagining how the Data Model would be presented in the generated app, given its separation from the GUI skeleton and the storyboard: ``it's not very clear how this will connect to another piece'' (P1). For this reason, participants suggested that Data Model could connect to views of the app where the Data Model appears, and users can click on UI content on each view to navigate to the corresponding Data Entity.

\subsubsection{GUI Skeletons}
On average, participants rated $3.63$ (SD = $0.98$) when asked about their agreement with the statement that they think \systemname helped them understand the generated code (between ``agree'' and `` strongly agree''), compared to $3.25$ (SD = $0.87$) when asked about the baseline. However, only 4 (33.3\%) participants preferred to use \systemname for the code generation step specifically, compared to 8 (66.7\%) that chose the baseline (\autoref{fig:comparisonResults}). This again is due to the slow speed in code generation as part of the limitation of the \systemname research prototype. 

\paragraph{Supporting user comprehension of UI code to be generated}

The GUI Skeletons helped participants understand the UI structure of each view to be generated in the app: ``you can relatively easily kind of see the structure and what each element is.'' (P2)
Specifically, the generated GUI skeletons helped them imagine how the code generated by the LLM would be structured, giving them more trust in the system, compared to the baseline condition where participants found the code generated by ChatGPT was like a ``black box'' (P5). Such transparency helped participants build mental models that effectively prepared them to work with the LLM's code generation output: 
``Prior to compiling code, I kind of understand like what the UI layout was and then prior to even getting the code, it was like kind of an agreement between me and the LLM, like this is how it's gonna flow, this is how it's gonna look like. So going into the code part, I could kind of like know where to look at.''

\paragraph{Allowing control on the UI code to be generated}
In the case where the GUI Skeletons suggested a UI structure that was different from what the developer imagined, \systemname allowed them to quickly debug before code generation: ``It throws in a bunch of some VStacks in it, I can quickly see based on the hierarchy, this is not gonna spit out good. It was lighter to do the [editing on] GUI skeleton and quicker.'' (P5) 
Because of this, participants were able to verify the generated app had the right structure before hitting the code generation button, saving time and effort:
``before I generate the code, I can make sure that everything's like in the right order and kind of what I wanted...when you add more like UI elements to it and more like modifiers, it gets, I feel like it could be more difficult to look for specific things. (P12)

\paragraph{Suggestions for GUI Skeleton}
At the same time, some more experienced iOS developers preferred to directly work with the generated code instead of editing the GUI skeleton: ``if I still had to make a lot of manual edits to the intermediate representation to achieve that, then I'd probably just prefer doing that in code instead.'' (P11) Some consider the GUI Skeleton a new syntax and would rather review the code directly, and thus hope to see a visually rendered preview of the skeleton.
Inline with this feedback, participants suggested a future version of \systemname could offer features that allow direct preview and manipulation on the UI to support iteration.%

Overall, the user study shows that \systemname supports developers in prototyping complex applications. Compared to the baseline, developers found the IRs provided by \systemname helpful in scaffolding understanding and exploration of the navigation flow at an early stage, getting them started with template UI content, and providing support in anticipating and controlling the code to be generated. With all these advantages and \systemname's ability to generate more complex applications, we did find that the code it generated tend to have more bugs. In the next section, we further dive into the type of bugs that \systemname generates, the effort to resolve them, and how that may impact the app development process.

\section{Technical Evaluation}
\label{sec:technical_eval}

\begin{table*}[t]
\centering
\caption{App size, number of compilation and navigation errors, and the effort required to resolve those errors (represented by the compilation and navigation diffs) across 10 SwiftUI apps generated by \systemname.}
\label{tab:error_resolution}
\begin{tabular}{lcccccc}
\toprule
       App & View Count & Lines of Code & Compilation Errors & Navigation Errors &  Compilation Diff & Navigation Diff \\
\midrule
   Pincast &          6 &           402 &                  0 &                 6 &              None &  4 files, +28, -15 \\
    TikTin &          6 &           419 &                  5 &                 6 &   3 files, +8, -8 &  3 files, +62, -43 \\
    AirBnP &          5 &           587 &                  6 &                 3 & 2 files, +10, -17 &   2 files, +12, -7 \\
    Mappin &          7 &           468 &                 24 &                 7 & 7 files, +43, -41 &  5 files, +96, -77 \\
   Slack'N &          5 &           529 &                 16 &                 3 & 5 files, +37, -41 &  3 files, +32, -18 \\
     Huber &          5 &           326 &                  4 &                 3 &   1 files, +5, -4 &  2 files, +39, -34 \\
   Shoppit &          5 &           515 &                  3 &                 3 &  3 files, +13, -6 &  2 files, +67, -48 \\
    FinBit &          6 &           521 &                  0 &                 1 &              None &     1 file, +8, -4 \\
Tintorship &         11 &           909 &                 13 &                 4 & 9 files, +24, -23 &  6 files, +65, -71 \\
    Zoomin &          8 &           472 &                  2 &                 4 &  1 files, +12, -2 & 7 files, +110, -63 \\
\midrule
\textbf{Averages} & 6.4 &         514.8 &                7.3 &               4.0 & 3.4 files, +16.9, -15.8 & 3.5 files, +51.9, -38.0  \\
\bottomrule
\end{tabular}
\end{table*}
\normalsize

The results of our user study show that participants preferred the baseline condition to \systemname for generating code. This is likely due to the higher number of compilation errors present in apps generated by our prototype (due to the larger volume of code in the more complex apps it produced). To better understand the \emph{types of errors} present in \systemname's output, we built several apps with \systemname and analyzed the errors in its output by their type, frequency, and difficulty to resolve.

The authors created 10 application concepts to prototype using \systemname by combining themes from different apps sampled from the Top 100 Free Apps Chart in the iOS App Store. For each app concept, an author created a prompt which described the high-level concept without requesting specific features. A list of these prompts is available in Appendix.

Each app prompt was sent to \systemname as an initial chat request. Once the storyboard was generated, one structural change was requested (e.g., a screen addition/removal) to simulate typical workflows. The resulting changes were used to generate intermediate representations and SwiftUI code, and then exported to Xcode for analysis.

For each application, we record the number of \emph{views}, \emph{lines of code}, and errors generated by \systemname. We distinguish between \emph{compilation errors}, which are emitted by the Swift compiler, and \emph{navigation errors}, which are logical errors in the implementation of the storyboard's navigation flows. An author with professional iOS development experience implemented fixes to these errors and recorded the diff size required for error resolution as a proxy for measuring the difficulty to resolve them.

We distinguish compilation and navigation errors because they occur at different semantic layers-—compilation errors arise from violations of formal language semantics, while navigation errors reflect failures in application-–level or design semantics. Moreover, compilation errors preclude application building, whereas navigation errors permit building but break conformance to the UI flow prescribed by the storyboard. Accordingly, resolution for each type demands different kinds and amount of engineering effort to bring the generated code to a functional state. Counts of each error per app are shown in \autoref{tab:error_resolution}, and detailed counts including types of each error are in Appendix.

\subsection{Compilation and Navigation Errors}
On average, generated apps had 7.3 compilation errors and 4.0 navigation errors.
Most of the compilation errors in the generated code were simple errors to resolve, resulting from the LLM hallucinating a property or code artifact which did not exist (e.g., invalid property access, invalid argument type). These errors were generally straightforward to correct with the help of IDE scaffolds like Fix-It~\cite{xcodeFixIt}. In future work, an agentic error correction approach would likely significantly reduce this type of error.

At least a few navigation errors—-where navigation flows were defined by the Storyboard but did not function at runtime—-were found in each app. The vast majority of these were placeholder comments the system inserted instead of implementing actual navigation code. Nonetheless, these errors were generally easy to resolve since navigation infrastructure was often present in affected views, but the logic to trigger navigation or link an outgoing view was missing.

The ``placeholder comment'' behavior emerged after we tried to suppress the insertion of ``TODO'' comments by updating the generation policy in system prompts.
One possible reason for this is that UI code is considered a low resource language~\cite{chenLowResource2022}, making up less than one percent of examples in some cases~\cite{muennighoff2023scalingdataconstrainedlanguagemodels}. SwiftUI and its navigation APIs (e.g., \texttt{NavigationStack}, \texttt{NavigationLink}) are relatively new, exacerbating their low resource nature.
This too might be addressable with a multi-pass agentic approach, where the model is asked to replace the comments with functioning code in a second pass.

\subsection{Compilation Error Outliers}
The average number of compilation error per application was 7. Three applications were significant outliers.

\subsubsection{Mappin}                  
The Mappin prototype allowed users to visually pin and share destinations on an interactive map. This app triggered 10 \textit{Missing Required Parameter} errors, all from invalid color literals, exceeding the total number of errors for all other applications. For reasons that are not entirely clear, the code generated by the LLM used custom colors, but each color was defined incorrectly. This may be due to the nature of UI code being a low-resource language.

\subsubsection{SlackN}
SlackN is an app that merges project management tools with instant chat messaging. The generated app had 6 \textit{Undeclared Identifier} compilation errors. In Swift, this occurs when a variable, function or type is referenced without being defined or available in the current scope. Most of these errors stemmed from the LLM hallucinating a variable or object that was not implemented. Manually implementing fixes for local variables was often simple, since the names provided enough context to infer their type and intent. Implementing similar fixes for objects posed a greater challenge because they were referenced in multiple parts of the app. Each object's structure had to be reverse–engineered from multiple call sites and the design intended by the LLM was not always obvious.

\subsubsection{Tintorship}
Tintorship is an app that matches mentors with mentees using a swipe-based interface. It was the most complex app generated in this evaluation, with 11 views and 909 lines of code, nearly double the average on both metrics. It surfaced 8 compilation errors, most of which were due to malformed member accesses in list constructions using SwiftUI. Lists were used generously in this app, and the model failed to ensure that each element in the list was uniquely identified, as SwiftUI requires. The cause of this is likely also due to UI code being a low-resource language.

\section{Discussion and Limitations}

Our goal in this paper was to explore the use of intermediate representations and an LLM to generate low-fidelity prototypes for functional multi-screen apps. 
We found the intermediate representations provided by \systemname scaffolds both the developer and the LLM in creating complex applications. Here we discuss our choice of intermediate representations and future directions for designing systems to support iterative scaffolded app generation with an LLM.

\paragraph{Choosing the right fidelity for intermediate representations.}
We specifically chose IRs that matched high-level steps in the app development process. Some of the more experienced user study participants expressed the desire to see code more early in the process, since their deep knowledge of the language could enable them to make modifications that could persist in code more quickly. This points to the question of designing more effective IRs for different usage scenarios, e.g., to help novice developers learn the SwiftUI language, or abstract code as much as possible to make app development with pure natural language possible, or getting to SwiftUI code earlier to support expert users.

\paragraph{Code LLMs have common failure modes.}
\systemname uses a powerful closed source LLM to generate code, however this LLM still makes mistakes which can lead to compiler errors when generating code. We have used many heuristics to clean this up, but \systemname is limited in the quality of code that it can produce by the underlying model that does the reasoning and generation. SwiftUI also is considered a low-resource UI language~\cite{wu-2024-uicoder}, and developing an LLM with more training in Swift and SwiftUI code could improve performance further.

\paragraph{Checkpointing and version control.}
While iterating with \systemname is much more fluid than switching context between design and development tools, deciding to revisit earlier stages in the app development process (e.g,. to add a new view to the wireframe) can still be costly as the intermediate steps may need to be inspected and modified before committing to code. One possibility to take advantage of LLMs to merge changes in earlier stages could be integrating version control~\cite{rawnVersionControl}, and mixing style components from various stages~\cite{yuwenMisty}.

\paragraph{Direct manipulation of the storyboard.}
While \systemname has been able to effectively produce and edit the app Storyboard through a chat interface, users may still prefer a direct manipulation interface for the Storyboard that updates the representation sent to the LLM. This is possible with more engineering effort.
Several participants wished they could also see rendered views in the Storyboard, which could possibly make the views proposed by the LLM easier to evaluate and intuit. We have experimented with this, and hope to add it to a future version of \systemname.

\paragraph{Threats to validity.}
We recruited all 12 participants from our organization, which might introduce organizational and social-desirability biases.

\section{Conclusion}
\label{section:conclusion}

Our work on \systemname shows the power of intermediate representations to scaffold design iteration between a user and an LLM, and demonstrates the feasibility of generating functional multi-view apps. While work remains to speed up generation and reduce bugs in the output, our user study showed that our participants preferred the \systemname prototype over a baseline for building an initial app idea from scratch. Our technical evaluation gives some insight into the challenges that current models face when generating code for complex UIs, which we believe can be solved in the future with agentic approaches. While \systemname remains a prototype, we hope that its ideas can be built on in future research to produce highly functional multi-screen UI authoring tools.

\bibliographystyle{ACM-Reference-Format}
\bibliography{sample-base}

\appendix
\onecolumn  %
\section{User Interface of \systemname}
\begin{figure}[!h]
    \centering
    \includegraphics[width=\textwidth]{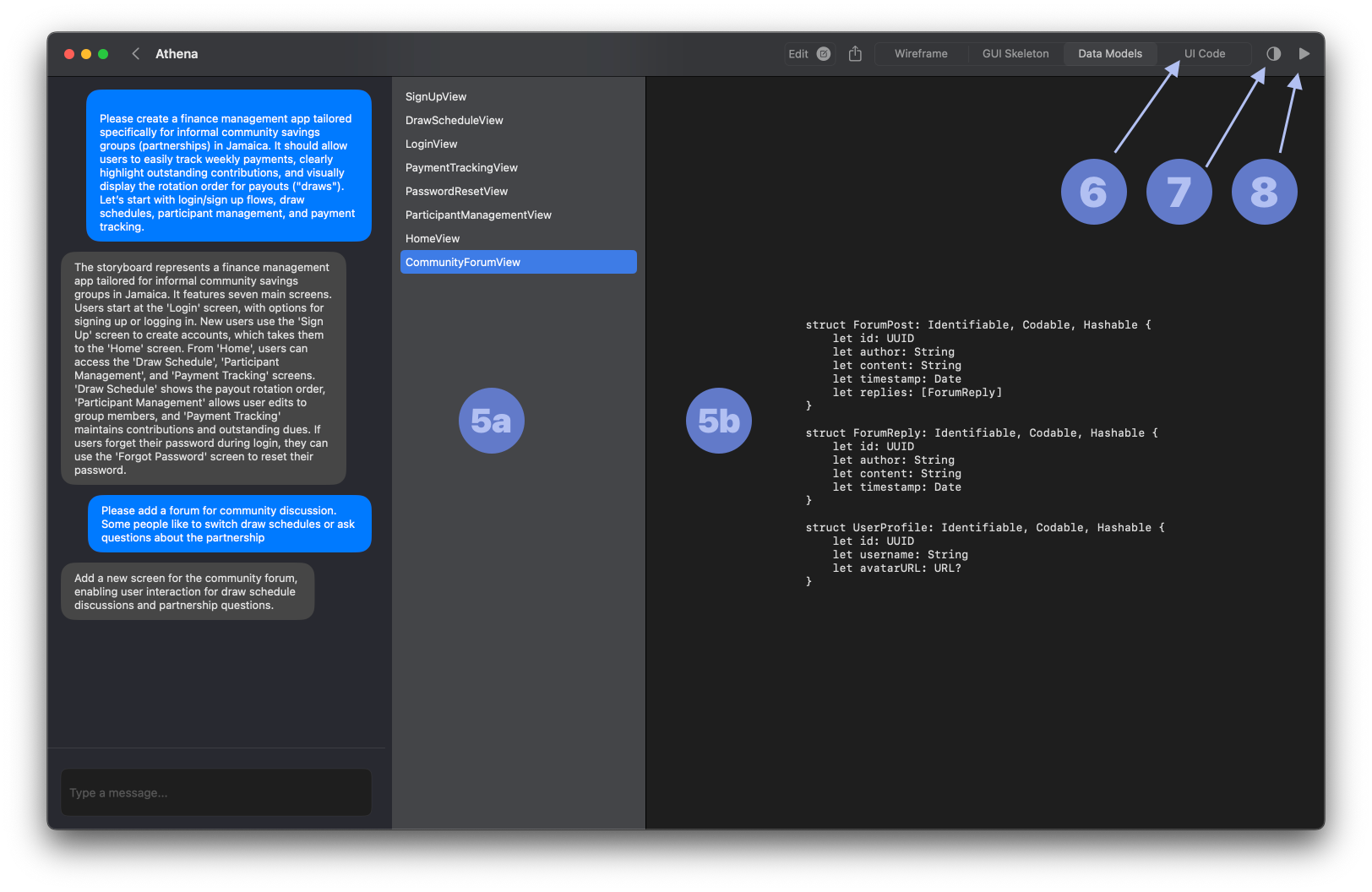}
    \caption{Screenshot of the \systemname user interface with the Data Model.}
    \label{fig:athenaUI.ir}
\end{figure}

The following figures show the user interface of \systemname. \autoref{fig:athenaUI.storyboard}: the Storyboard view with ha chat interface to propose the app idea and request changes (1, 2). The generated Storyboard is visible on the right (3). \autoref{fig:athenaUI.ir}: the Data Model editor view showing the list of views corresponding to screens in the Storyboard (5a) and the content of the selected view for the Data Model (5b). Users can switch to the UI Skeleton editor view using the UI Skeleton tab, and the interface is the same as that of Data Model. Data Model and UI Skeleton are generated for a Storyboard once (7) is clicked, and full SwiftUI code is generated which instentiates the IRs with (8). 

\begin{figure}[!h]
    \centering
    \includegraphics[width=\textwidth]{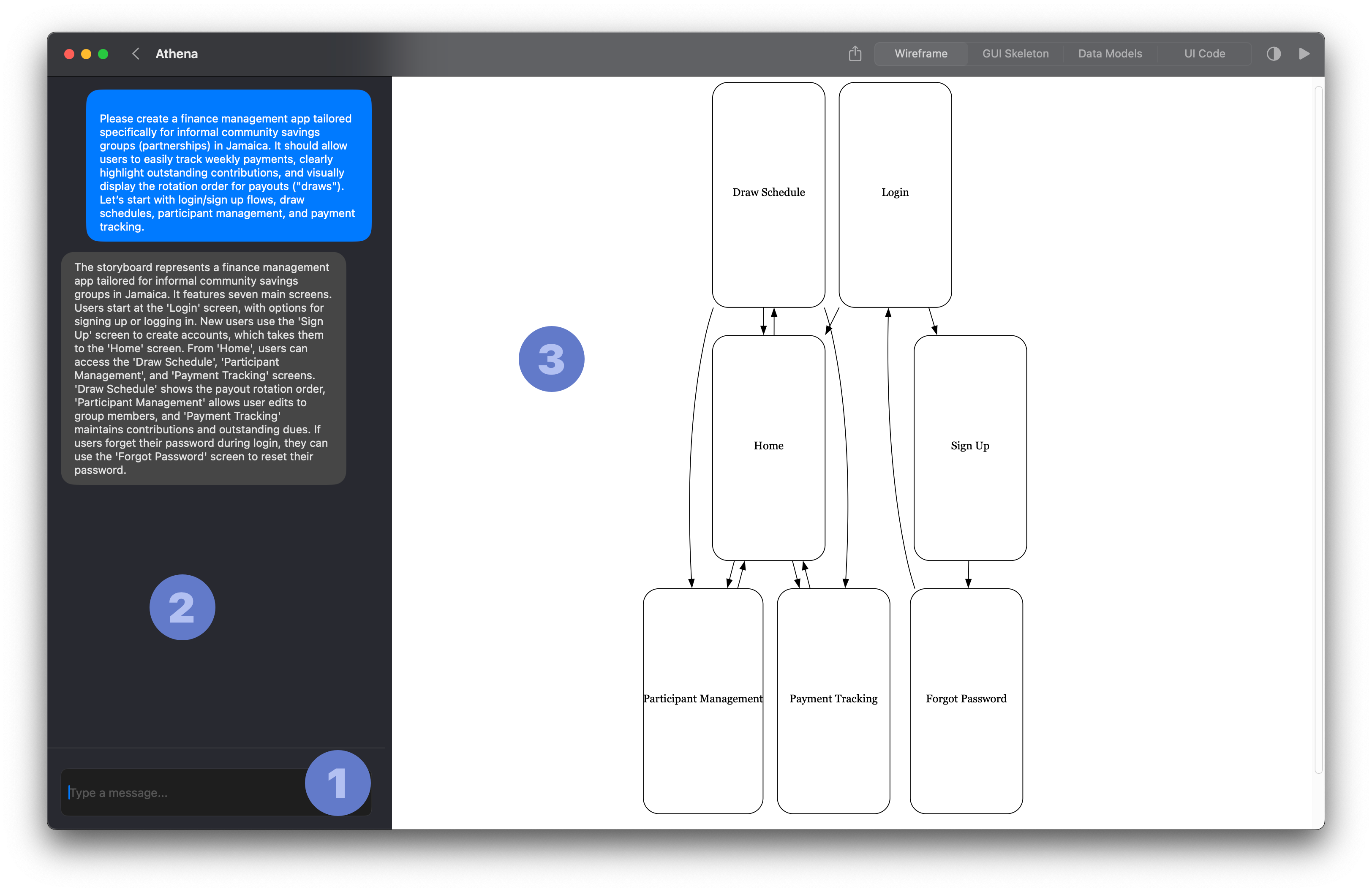}
    \caption{Screenshot of the \systemname user interface with the Storyboard.}
    \label{fig:athenaUI.storyboard}
\end{figure}

\clearpage
\section{Intermediate Representation Examples}
\label{sec:irExamples}

The following figures (\autoref{fig:suppWireframe} to \autoref{supp:planResponse}) present examples for the intermediate representations. 

\begin{figure*}[h!]
    \centering
    \small
    \begin{lstlisting}[language=json]
{
  "storyboard": {
    "description": "Fashion-forward shoe marketplace app.",
    "nodes": [
      {
        "id": 1,
        "name": "Home",
        "description": "This is the home screen...",
        "swiftUIViewName": "HomeView",
        "outgoingEdges": [
          2
        ]
      },
      {
        "id": 2,
        "name": "Product Detail",
        "description": "This screen shows...",
        "swiftUIViewName": "ProductDetailView",
        "outgoingEdges": [
          1
        ]
      }
    ]
  },
  "explanation": "The storyboard represents..."
}\end{lstlisting}
    \caption{Abbreviated wireframe}
    \label{fig:suppWireframe}
\end{figure*}

\begin{figure*}[h!]
    \centering
    \begin{lstlisting}[language=json]
{
  "views": [
    {
      "id": 2,
      "name": "Product Detail",
      "swiftUIViewName": "ProductDetailView",
      "transitions": [
        {
          "destination": "WriteReviewView",
          "type": "sheet",
          "trigger": "onWriteReviewButtonTap",
          "dataPass": {
            "items": [
              "product.id",
              "product.name"
            ]
          }
        }
      ]
    }
  ]
}\end{lstlisting}
    \caption{Abbreviated modification plan}
    \label{fig:suppPlan}
\end{figure*}

\begin{figure*}[h!]
    \centering
    \begin{lstlisting}[language=json]
{
  "viewName": "NotesListView",
  "id": 1,
  "guiSkeleton": {
    "StateVariables": ["notes"],
    "Layout": {
      "MainContainer": {
        "Elements": [
          {
            "List": {
              "DataSource": "notes",
              "Elements": [
                {
                  "HStack": {
                    "Elements": [
                      { "Text": { "Value": "note.title" } },
                      { 
                          "Button": { 
                            "Label": "Edit",
                            "OnTap": {  "Navigate": { "Destination": "EditNoteView" }}
                        } 
                     }
                    ]
                  }
                }
              ]
            }
          },
          {
            "Button": { 
                "Label": "Add Note",
                "OnTap": {
                    "Navigate": {
                        "Destination": "AddNoteView"
                    }
                } 
            }
          }
        ]
      },
      "Navigation": { "NavigationBar": { "Title": "My Notes" } }
    }
  }
}\end{lstlisting}
    \caption{Abbreviated GUI Skeleton}
    \label{fig:suppGuiSkeleton}
\end{figure*}

\begin{figure*}[h!]
    \begin{lstlisting}[language=json]
{
  "colors": {
    "primary": "#F0F0F0",
    "secondary": "#1DB954",
    "accent": "#FF5733",
    "neutral": {
      "dark": "#333333",
      "medium": "#777777",
      "light": "#BBBBBB"
    }
  },
  "typography": {
    "font": "SF Pro Display",
    "h1": { "weight": "Bold", "size": 36 },
    "h2": { "weight": "SemiBold", "size": 28 },
    "body": { "weight": "Regular", "size": 16 },
    "caption": { "weight": "Regular", "size": 12 }
  },
  "components": {
    "button": {
      "standard": { "bgColor": "#1DB954", "textColor": "#FFFFFF", "radius": 8 },
      "primary": { "bgColor": "#FF5733", "textColor": "#FFFFFF", "radius": 12 }
    },
    "navBar": {
      "bgColor": "#F0F0F0",
      "title": { "fontSize": 18, "color": "#333333" },
      "button": { "fontSize": 16, "color": "#1DB954" }
    },
    "tabBar": {
      "bgColor": "#F0F0F0",
      "iconColor": "#777777",
      "selectedColor": "#1DB954",
      "labelFontSize": 12
    },
    "card": {
      "bgColor": "#FFFFFF",
      "radius": 10,
      "shadow": { "color": "#000000", "opacity": 0.1, "offsetY": 4, "blur": 6 }
    }
  },
  "icons": {
    "style": "Minimal line icons",
    "sizes": [24, 32],
    "system": ["play.circle", "pause.circle", "gear", "music.note"]
  },
  "animations": {
    "duration": "0.25s",
    "easing": "ease-in-out",
    "style": "slide screens, fade content"
  }
}\end{lstlisting}
    \caption{Abbreviated application design system}
    \label{fig:suppAppDesignSystem}
\end{figure*}

\begin{figure*}[h!]
    
    \begin{lstlisting}[language=json]
{
  "purpose": "Allow users to log in to access app features.",
  "layout": "Centered form: email/password fields, login button, 'Forgot password?' link.",
  "interactions": {
    "gestures": "Tap inputs/buttons, swipe down to dismiss keyboard.",
    "feedback": "Buttons highlight with subtle vibration; input fields show focus.",
    "keyboard": "Fields remain visible above keyboard."
  },
  "navigation": {
    "entryPoint": "Initial screen; successful login goes to Home."
  },
  "actions": {
    "primary": "Log in",
    "secondary": "Password recovery"
  },
  "visual": {
    "colors": {
      "background": "#F0F0F0",
      "inputs": "#FFFFFF (light border)",
      "button": "#1DB954",
      "text": "#333333"
    },
    "typography": {
      "heading": "Bold, 28pt",
      "inputs": "Regular, 16pt",
      "button": "SemiBold, 16pt"
    },
    "animations": "Inputs animate on focus; button press animation."
  },
  "inputs": {
    "style": "Rounded minimal, gray placeholder.",
    "validation": "Real-time email/password checks."
  },
  "errors": {
    "message": "Invalid credentials. Try again.",
    "visual": "Red border, message below fields."
  },
  "loading": {
    "indicator": "Spinner on login button during processing."
  }
}\end{lstlisting}
    \caption{View design system for a login view}
\end{figure*}

\begin{figure*}[h!]
    \begin{lstlisting}[language=json]
{
  "views": [
    {
      "id": 1,
      "name": "Home",
      "swiftUIViewName": "HomeView",
      "viewCode": "// Complete SwiftUI implementation of HomeView..."
    },
    {
      "id": 2,
      "name": "ProductDetail",
      "swiftUIViewName": "ProductDetailView",
      "viewCode": "// Complete SwiftUI implementation of ProductDetailView..."
    },
    {
      "id": 0,
      "name": "Purchase",
      "swiftUIViewName": "PurchaseView",
      "viewCode": "// Complete SwiftUI implementation of PurchaseView"
    }
  ],
  "utilities": [
    {
      "name": "Color+Extension",
      "code": "// Complete implementation of Color+Extension..."
    }
  ]
}\end{lstlisting}
    \caption{Abbreviated SwiftUI view generation output}
    \label{supp:viewGeneration}
\end{figure*}

\begin{figure*}[h!]
    \begin{lstlisting}[language=json]
{
  "changeType": "guiSkeleton",
  "storyboardChanges": {
    "addScreens": [{"id":101,"name":"UserProfileView"}],
    "removeScreens": [{"id":50,"name":"OldSettingsView"}],
    "addConnections": [{"from":101,"to":102}],
    "removeConnections": [{"from":50,"to":51}]
  },
  "guiSkeletonChanges": {
    "filesToModify": [{"swiftUIViewName":"UserProfileView","id":101}],
    "newFilesToCreate": [{"swiftUIViewName":"UserDetailsView","id":102}],
    "filesToDelete": [{"swiftUIViewName":"OldSettingsView","id":50}]
  },
  "dataModelChanges": {
    "filesToModify": [
      {"swiftUIViewName":"UserModel","id":201},
      {"swiftUIViewName":"UserService","id":202}
    ]
  },
  "technicalDescription": {
    "summary": "Added user age support; Removed OldSettingsView."
  }
}\end{lstlisting}
    \caption{Abbreviated plan response}
    \label{supp:planResponse}
\end{figure*}

\clearpage  %
\twocolumn  %

\newpage
\section{User Study Details}
\label{sec:userStudyAppDetails}

\subsection{User Study Participants}
\label{sec:studyParticipantsDetail}

\autoref{tab:participantDetail} shows the profile and the order and conditions for each participants in the user study. The conditions are explained as follows: 
\begin{enumerate}
\item AM: \systemname for the Music Task
\item BP: baseline for the Parking Task
\item BM: baseline for the Music Task
\item AP: \systemname for the Parking Task

\end{enumerate}

\begin{table}[h] 
    \caption{User study participants profile}
    \label{tab:participantDetail}
    \begin{tabular}{lcccc}
    \toprule
    ID  & Gender & Job title          & YOE w/ iOS & Condition \\
    \midrule
    P1  & M      & Research Scientist &  2-5                        & AM, BP    \\
    P2  & M      & Software Engineer  & \textgreater{}10         & AM, BP    \\
    P3  & M      & Software Engineer  & 2-5                      & AP, BM    \\
    P4  & M      & Research Engineer  & 2-5                      & BM, AP    \\
    P5  & M      & Software Engineer  & 1-2                      & BP, AM    \\
    P6  & M      & Software Engineer  & \textgreater{}10         & AM, BP    \\
    P7  & F      & Research Scientist & 2-5                      & AP, BM    \\
    P8  & M      & Software Engineer  & \textgreater{}10         & BP, AM    \\
    P9  & M      & Software Engineer  & \textgreater{}10         & BM, AP    \\
    P10 & F      & Software Engineer  & 5-10                     & BM, AP    \\
    P11 & M      & Research Engineer  & 1-2                         & AP, BM    \\
    P12 & F       & Software Engineer                   & 1-2                         & BP, AM   \\
    \bottomrule
    \end{tabular}
    
\end{table}

\subsection{User Study Questions}
\label{sec:userStudyQuestionsDetail}

For the \systemname condition, the agreement statements for participants to rate from 1 (strongly disagree) to 5 (strongly agree) include: 

\begin{itemize}
    \item ``The generated storyboard helps me think about the navigation flow that I did not think about before;'' 
    \item ``The system is helpful for me to iterate on the navigation flow of the app;'' 
    \item ``The data model is helpful for me to understand what content will be presented on each view;'' 
    \item ``The system is helpful for me to iterate on the content that will be presented on each view;'' 
    \item ``The system helps me understand how \systemname will implement the app before exporting;'' 
    \item ``The system is helpful for me to iterate on how the app will be implemented.'' 
\end{itemize}

For the baseline condition, the statements include:

\begin{itemize}
    \item ``ChatGPT is helpful for me to iterate on the navigation flow of the app;'' 
    \item ``ChatGPT is helpful for me to iterate on the content that will be presented in each view;'' 
    \item ``ChatGPT is helpful for me to iterate on how the app will be implemented.'' 
\end{itemize}

\subsection{Apps Generated in the User Study}

\autoref{tab:baseline.metrics} and \autoref{tab:app.metrics} present the number of views and lines of code in the apps generated with the baseline system and \systemname in the user study. 

\begin{table}[!ht]
    \centering
    \caption{App size generated by baseline condition by each participant.}
    \label{tab:baseline.metrics}
    \begin{tabular}{lrr}
    \toprule
    \textbf{App} & \textbf{View Count} & \textbf{Lines of Code} \\
    \midrule
    P1  & 2 & 101 \\
    P2  & 3 & 139 \\
    P3  & 3 & 123 \\
    P4  & 2 & 72  \\
    P5  & 5 & 150 \\
    P6  & 2 & 77  \\
    P7  & 2 & 62  \\
    P8  & 4 & 138 \\
    P9  & 6 & 149 \\
    P10 & 4 & 176 \\
    P11 & 4 & 121 \\
    P12 & 4 & 120 \\
    \bottomrule
    \end{tabular}
    
\end{table}

\begin{table}[!ht]
    \centering
    \caption{App size generated in \systemname by each participant.}
    \label{tab:app.metrics}
    \begin{tabular}{lrr}
    \toprule
    \textbf{App} & \textbf{View Count} & \textbf{Lines of Code} \\
    \midrule
    P1  & 4  & 290 \\
    P2  & 5  & 357 \\
    P3  & 10 & 382 \\
    P4  & 7  & 424 \\
    P5  & 5  & 240 \\
    P6  & 8  & 525 \\
    P7  & 3  & 221 \\
    P8  & 4  & 259 \\
    P9  & 7  & 405 \\
    P10 & 7  & 368 \\
    P11 & 4  & 358 \\
    P12 & 6  & 461 \\
    \bottomrule
    \end{tabular}
\end{table}

\clearpage
\onecolumn

\section{Technical Evaluation Details}

\autoref{tab:navigation_errors} and \autoref{tab:compilation_errors} show the types of navigation errors and compilation errors respectively from the experiments reported in the \autoref{sec:technical_eval}. 

\autoref{tab:appCreationPrompts} presents the prompts used in creating the example apps used in \autoref{sec:technical_eval}.

\begin{table}[h!]
    \centering
    \caption{Initial app creation prompts and change request prompts used for apps in technical evaluation (\autoref{sec:technical_eval})}
    \label{tab:appCreationPrompts}
\begin{tabular}{p{0.08\linewidth}p{0.5\linewidth}p{0.32\linewidth}}
\toprule
\textbf{App}        & \textbf{Prompt}                                                                                                                                                                                                                                                                                                         & \textbf{Change}                                                                                                                                                           \\ \midrule
\textbf{Pincast}    & Create podcast discovery and curation app that takes inspiration from Pinterest's visual layout and social features, but focuses entirely on audio content.                                                                                                                                                             & Add a signup/login flow and connect it to the home view                                                                                                                   \\ \midrule
\textbf{TikTin}     & Create a professional networking app where users share short, engaging video content showcasing their expertise, projects, or company culture. This app combines LinkedIn's professional focus with TikTok's video format to foster authentic connections and knowledge sharing among professionals.                    & Make messaging accessible from the home view. Remove wlecome, recommendations, notifications, Edit MetaData, Discover and search, and AR tools.                           \\ \midrule
\textbf{AirBnP}     & An app that lets people rent out unused parking spots to drivers, combining Airbnb’s peer-to-peer model with real-time availability maps. This approach eases urban parking congestion while helping spot owners earn extra income.                                                                                     & Remove the connection from booking confirmation to Home                                                                                                                   \\ \midrule
\textbf{Mappin}     & Create an app where users can visually pin and share their favorite destinations on an interactive map. This app combines Pinterest's visual inspiration with Google Maps' functionality, enabling users to create and share personalized travel itineraries and discover new places through community-curated content. & Remove 'Pin Destination and add a login view connected to the map view                                                                                                    \\ \midrule
\textbf{Slack’N}    & Build an app that merges Slack's real-time communication with Notion's project management capabilities. Team members can manage tasks and share updates                                                                                                                                                                 & There should be no connection between project detail and home, remove 'Inbox', and there should be no connection between 'TaskCreate and home                             \\ \midrule
\textbf{Huber}      & Create an app that connects patients with nearby healthcare professionals for immediate consultations or home visits. Applying Uber's real-time matching and tracking technology to healthcare, the app offers convenient access to medical services when and where they're needed.                                     & Remove the landing screen                                                                                                                                                 \\ \midrule
\textbf{Shoppit}    & Build an e-commerce app where product listings are enhanced with community discussions, upvotes, and reviews. This approach combines Amazon's marketplace with Reddit's user engagement, helping shoppers make informed decisions through collective insights.                                                          & Remove community                                                                                                                                                          \\ \midrule
\textbf{FinBit}     & Build an app that tracks users' spending habits and financial health in real-time, similar to how Fitbit monitors physical activity. With interactive dashboards and goal-setting features, it encourages better money management and financial awareness.                                                              & Remove notifications and add a login screen                                                                                                                               \\ \midrule
\textbf{Tintorship} & Build an app that matches mentors and mentees using a swipe-based interface. By combining Tinder's matching system with professional profiles, it facilitate meaningful career development connections based on shared interests and goals.                                                                             & Remove the chat screen-messaging alone is fine                                                                                                                            \\ \midrule
\textbf{Zoomin}     & Create an online education app that blends Coursera's structured courses with Zoom's live video capabilities. This allows for interactive classes with real-time instructor feedback.                                                                                                                                   & Remove course content, course detail is enough. You should be able to profile from course dashboard. You should be able to get to the discussion from from course detail. \\ \midrule
\end{tabular}
\end{table}

\clearpage

\label{sec:techEvalErrorDetails}

\begin{table*}[t]
\centering
\caption{Navigation Error Frequency}
\label{tab:navigation_errors}
\begin{tabular}{lcccccccccccc}
\toprule
              Error Type & Tintorship & FinBit & Shoppit & Mappin & Huber & Slack'N & Zoomin & Pincast & TikTin & AirBnP & Total \\
\midrule
 Missing Navigation Link &          0 &      0 &       0 &      0 &     0 &       1 &      1 &       2 &      2 &      0 &     6 \\
      Navigation Comment &          1 &      1 &       3 &      5 &     3 &       1 &      1 &       4 &      3 &      0 &    22 \\
Navigation Closure Empty &          0 &      0 &       0 &      0 &     0 &       0 &      0 &       0 &      0 &      0 &     0 \\
 Missing Navigation View &          0 &      0 &       0 &      0 &     0 &       0 &      0 &       0 &      1 &      0 &     1 \\
              API Misuse &          0 &      0 &       0 &      0 &     0 &       0 &      1 &       0 &      0 &      2 &     3 \\
     No Navigation Logic &          2 &      0 &       0 &      2 &     0 &       1 &      1 &       0 &      0 &      1 &     7 \\
  Wrong Destination View &          1 &      0 &       0 &      0 &     0 &       0 &      0 &       0 &      0 &      0 &     1 \\
                   Total &          4 &      1 &       3 &      7 &     3 &       3 &      4 &       6 &      6 &      3 &    40 \\
\bottomrule
\end{tabular}

\bigskip
\begin{minipage}{\textwidth}
\centering
\small
Navigation error counts aggregated over 10 generated SwiftUI apps. Compilation errors are shown in Table~\ref{tab:compilation_errors}.
\end{minipage}

\end{table*}

\normalsize

\begin{table*}[t]
\centering
\caption{Compilation Error Frequency}
\label{tab:compilation_errors}
\begin{tabular}{lccccccccccc}
\toprule
Error Type & Tintorship & FinBit & Shoppit & Mappin & Huber & Slack'N & Zoomin & Pincast & TikTin & AirBnP & Total \\
\midrule
Missing Required Parameter & 0 & 0 & 0 & 10 & 0 & 1 & 0 & 0 & 2 & 0 & 13 \\
Invalid Property Access & 1 & 0 & 1 & 4 & 1 & 2 & 0 & 0 & 3 & 0 & 12 \\
Invalid Argument Type & 2 & 0 & 0 & 2 & 0 & 3 & 0 & 0 & 0 & 4 & 11 \\
Immutability Violation & 2 & 0 & 0 & 0 & 1 & 0 & 0 & 0 & 0 & 1 & 4 \\
Protocol Conformance Error & 2 & 0 & 0 & 1 & 0 & 1 & 0 & 0 & 0 & 0 & 4 \\
Missing Import & 0 & 0 & 0 & 1 & 0 & 0 & 1 & 0 & 0 & 0 & 2 \\
Malformed Member Access & 5 & 0 & 1 & 0 & 0 & 0 & 0 & 0 & 0 & 0 & 6 \\
Access Control Violations & 0 & 0 & 0 & 5 & 0 & 2 & 0 & 0 & 0 & 0 & 7 \\
Undeclared Identifier & 0 & 0 & 1 & 1 & 0 & 6 & 1 & 0 & 0 & 0 & 9 \\
Type Usage Violation & 0 & 0 & 0 & 0 & 1 & 1 & 0 & 0 & 0 & 0 & 2 \\
Generic Inference Failure & 0 & 0 & 0 & 0 & 1 & 0 & 0 & 0 & 0 & 0 & 1 \\
Invalid Parameter Usage & 1 & 0 & 0 & 0 & 0 & 0 & 0 & 0 & 0 & 1 & 2 \\
Total & 13 & 0 & 3 & 24 & 4 & 16 & 2 & 0 & 5 & 6 & 73 \\
\bottomrule
\end{tabular}

\bigskip
\begin{minipage}{\textwidth}
\centering
\small
Compilation error counts aggregated over 10 generated SwiftUI apps. Navigation errors are shown in Table~\ref{tab:navigation_errors}.
\end{minipage}

\end{table*}

\normalsize

\end{document}